\documentclass[12pt]{article}
\pdfoutput=1
\usepackage{amsmath}
\usepackage{amssymb}
\usepackage{feynmp}
\usepackage{graphicx}
\usepackage{epstopdf}
\usepackage{bm}
\usepackage{simplewick}
\usepackage[font=small]{caption,subfig}

\newcommand\be{\begin{eqnarray}}
\newcommand\ee{\end{eqnarray}}
\newcommand\we{\wedge}
\newcommand\R{\mathbb{R}}
\newcommand\C{\mathbb{C}}
\newcommand\x{\times}
\renewcommand\.{\cdot}
\newcommand\vf{\varphi}
\renewcommand\a{\alpha}
\renewcommand\b{\beta}
\newcommand\de{\delta}
\newcommand\g{\gamma}
\newcommand\e{\epsilon}
\newcommand\ka{\kappa}
\newcommand\De{\Delta}
\newcommand\w{\omega}
\newcommand   \pdr  [2] {\frac{\partial #1}{\partial #2}}
\newcommand   \lr  [3] {\left #1 #2 \right #3}
\newcommand   \p   [1] {\lr({#1})}
\newcommand\ptl{\partial}
\newcommand\half{\frac 1 2}
\newcommand\inv{^{-1}}
\newcommand\aeq{\approx}
\newcommand\Tr{\mathrm{Tr}}
\newcommand\KK{\mathrm{KK}}
\newcommand\cM{\mathcal{M}}
\newcommand\bF{\mathbf{F}}
\newcommand\bA{\mathbf{A}}
\newcommand\Labc{Q_{a,0}^{b,c}}
\newcommand\Labcdef{Q_{a,b}^{c+e,d+f}}
\newcommand\Laab{Q_{a,0}^{a,b}}
\newcommand\ad{\mathrm{Ad}}
\newcommand\GUT{\mathrm{GUT}}
\newcommand\eV{\,\mathrm{eV}}
\renewcommand\bar[1]{\overline{#1}}
\newcommand\tl[1]{\widetilde{#1}}
\newcommand\Lflxa{Q_{a,0}^{1,0}}
\newcommand\Lflxb{Q_{a,0}^{2,0}\dots}
\newcommand\Rflxa{D_{0,b}^{0,1}}
\newcommand\Rflxb{\dots}
\newcommand\Hone{H^{1,0}}
\newcommand\Htwo{H^{2,0}\dots}
\newcommand\Hthree{H^{a,1}}
\newcommand\Hfour{H^{a,2}\dots}

\long\def\symbolfootnote[#1]#2{\begingroup%
\def\thefootnote{\fnsymbol{footnote}}\footnote[#1]{#2}\endgroup}

\begin{document}

\begin{titlepage}
\noindent
\vspace{2cm}

\begin{center}
  \begin{Large}
    \begin{bf}
Quark and Lepton Flavor Physics from F-Theory\\
     \end{bf}
  \end{Large}
\end{center}
\vspace{0.2cm}

\begin{center}

\begin{large}
Lisa Randall\symbolfootnote[1]{randall@physics.harvard.edu} and David Simmons-Duffin\symbolfootnote[2]{davidsd@physics.harvard.edu}\\
\end{large}
\vspace{0.3cm}
  \begin{it}
Jefferson Physical Laboratory, Harvard University,\\
Cambridge, Massachusetts 02138, USA
\vspace{0.5cm}
\end{it}\\

\end{center}

\center{\today}

\begin{abstract}
Recent work on local F-theory models shows the potential for new
categories of flavor models. In this paper we investigate the
perturbative effective theory interpretation of this result. We also
show how to extend the model to the neutrino sector.
\end{abstract}

\end{titlepage}

\setcounter{page}{2}

\section{Introduction} The well-known flavor problem has perplexed particle theorists for quite some time now.
With each new measurement the  mysteries seem to increase.  For
example, measurements of neutrino masses to date have defied
expectations, with large mixing angles and at least two of the
neutrinos significantly more degenerate than in the quark sector.
Turned around, the patterns of masses and mixings in the quark and
lepton sector could be providing some important clues about
underlying physics -- physics that admittedly is often hard to test.

Of course there are many possible flavor models at this point, and it
is difficult to distinguish among them. Only a few are elegant
enough to avoid requiring many new arbitrary charges or parameters.
 In this sense higher-dimensional models seem promising, in that wavefunctions can naturally account for
hierarchies and angles \cite{Grossman:1999ra,Gherghetta:2000qt,Fitzpatrick:2007sa,Perez:2008ee}. Also many predictions in the end rely only on the nature of the wavefunctions in higher-dimensional space. The particular types of wavefunctions and Yukawas for the models discussed in \cite{Heckman:2008qa} and in this paper use the existence of models based on 7-branes in a significant way, as we will discuss and expand on.

In this paper we consider the recent F-theory models of flavor
proposed in \cite{Heckman:2008qa,Beasley:2008dc,Beasley:2008kw}. We
try to identify the distinguishing features that might make these
models special. We show the models do have a low-energy effective
field theory interpretation in a Froggett-Nielsen-like form but that
to truly reproduce the F-theory predictions would require a model
that looks inelegant from a low-energy point of view. The presence
of KK modes in the higher-dimensional theory automatically provides
extra flavor-carrying states whose presence influences the structure
of the low-energy mass matrices.

We also point out the qualitative features we find most important about these models.
One is that they predict approximately rank-one matrices for both up and down quark Yukawas, a prediction that
 seems supported by what we know. Furthermore we argue that the most natural implementation of neutrinos would
 involve an approximately rank-two matrix, predicting  a third small eigenvalue. Finally we argue that the
framework
 is sufficiently general to support the measured mixing angles, though we find predictions of these parameters
less robust,
 but completely compatible with known numbers.

\section{Review of F-theory GUTs}

We now review the set-up of Refs.
\cite{Heckman:2008qa,Beasley:2008dc,Beasley:2008kw}.  At and below the GUT scale, an F-theory GUT is a higher dimensional
QFT whose degrees of freedom live on submanifolds of a complex $3$-fold $B_3$
(times $\R^{1,3}$). Gravity lives in the bulk, and we'll assume the
decoupling limit of \cite{Beasley:2008dc}, where $M_{pl}\to\infty$.
$7$-branes with world-volume $\mathcal N=1$ SUSY gauge theories are
wrapped on complex surfaces\footnote{2-complex dimensional, or
4-real dimensional.} $S,S',\ldots\subset B_3$.  The associated gauge
groups $G_S, G_{S'},\dots$ depend on the number and type of
$7$-branes wrapping each surface.\footnote{More precisely, in
F-theory, the 3-fold $B_3$, the surfaces $S,S'$, and the types of
branes wrapping those surfaces are all encoded in the geometric data
of a compact elliptically fibered Calabi-Yau 4-fold.  For instance,
$7$-branes wrap loci where the elliptic fiber of the 4-fold
degenerates, and the associated gauge group is specified by the
singularity type of the fibration along the brane.  In this paper,
we won't need most of these details, and it will suffice to take the
7-branes and their associated field theories (derived in \cite{Beasley:2008dc}) as an effective
description.}    Matter resides in 1-complex dimensional
intersections of two such surfaces, $\Sigma=S\cap S'$, called
``matter curves."

\begin{figure}[!ht]
\centering
\includegraphics[width=.5\textwidth]{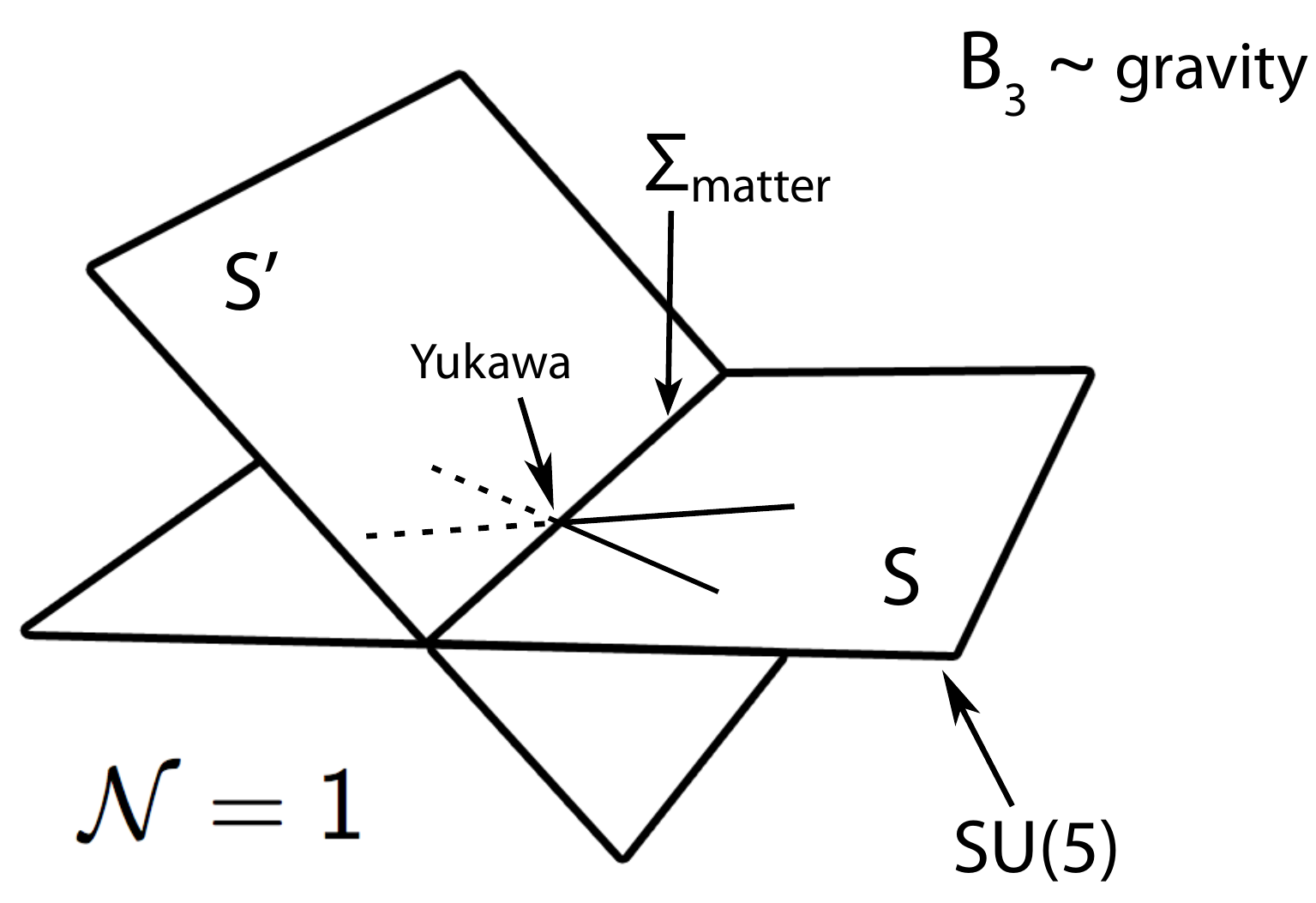}
\\
\vspace{.4in}
\begin{tabular}{|l|l|l|}
\hline
dim. & internal dim. & feature \\
\hline
10 & $6=\dim(B_3)$ & gravity\\
8 & $4=\dim(S)$ & gauge fields\\
6 & $2=\dim(S\cap S')$ & matter\\
4 & $0=\dim(S\cap S'\cap S'')$ & interactions\\
\hline
\end{tabular}
\caption{The structure of an F-theory GUT}
\end{figure}

We can identify the matter localized on an intersection of two
branes $\Sigma\subset S\cap S'$ by beginning with both branes on top
of each other (which has a simple description as a ``parent"
8-dimensional effective gauge theory), and then rotating $S'$ off of
$S$ by turning on a linear vev for the field $\vf$ representing the
transverse distance between the branes. Degrees of freedom trapped
near $\vf=0$ make up the theory in the matter curve $S\cap S'$. This
Riemann surface supports vector-like matter charged under $G_S\x
G_{S'}$.

A critical distinction for these models is that several distinct
flavors can live on a single such matter curve. An index theorem
determines the number of massless modes that reside on  a Riemann
surface formed by the intersection of $S$ and $S'$. This allows for
the possibility of three zero-mode solutions on these surfaces, each
with its own independent wavefunction. This is not generally the
case for models based on brane intersections, where generally
multiple generations exist on the same surface only when multiple
branes coincide, in which case the  three generations would have the
same gauge charges and wavefunctions, making hierarchies difficult
to establish geometrically. In the F-theory set-up, the three
generations can participate differently in the Yukawas, and this is
in fact generally the case. This is the critical feature that allows
for an interesting geometrical generation of the mass hierarchies
and mixing angles.

The Yukawa couplings responsible for masses come from superpotential interactions
between matter fields arising on point-like intersections of the associated matter curves.
Because these matter representations are the residue of ``parent" gauge interactions
that are broken when the the 7-branes do not coincide, the Yukawa interactions can be thought of as arising
from gauge interactions and supersymmetry. Because the intersecting branes are in different orientations,
 the associated gauge bosons are heavy, so the Yukawas survive without additional spurious gauge interactions.

In the situations considered in
\cite{Heckman:2008qa,Beasley:2008dc,Beasley:2008kw}, the geometry of
the GUT brane $S$ precludes the existence of bulk zero modes resembling SM matter. Therefore SM-like quark Yukawas require three
additional surfaces to intersect $S$ at a single point. We note here
that although three surfaces intersecting is generic, four surfaces
is not and we view this as an additional assumption. However, if you
view an exceptional gauge group in the parent theory as fundamental,
and the surfaces as arising from breaking the gauge symmetry, the
Yukawa and four-way intersection is automatic \cite{Beasley:2008kw}.
This is entirely possible in the local description but it is
ultimately important to see if this assumption will be realized in a
full global model \cite{FutureMarsano}.

At this point we already see that Yukawa matrices are approximately
rank-one, since the intersection of the three matter curves on $S$
occurs only at a single point in the internal dimensions, and the
Yukawa is roughly the outer product of the three left- and right-handed wavefunction vectors. This goes a long way toward realizing
the structure we see in the quark masses.  We will soon see that
with the true wavefunctions, one finds the hierarchical mass
matrices we know to describe the quark sector.

\subsection{Matter Zero Mode Wavefunctions}

It will be important for us to understand in more detail how SM matter arises when branes are rotated
apart starting from a parent gauge theory, and what determines the zero mode wavefunctions.
We'll quickly review the discussion in \cite{Beasley:2008dc}, itself a review of \cite{Katz:1996xe}.

Suppose that $S$ has some GUT group
$G_S$ (which for concreteness we can take to be SU(5)), $S'$ has gauge group $U(1)$, and the parent theory has gauge
group $G$, broken to $G_S\x U(1)$ by turning on a generator $T\in
\ad_G$.  The degrees of freedom in the parent gauge theory (with
$\mathcal N=1$ superpartners paired up) are \be
A_\mu, \eta_\a && \mbox{scalars on $S$, in $\ad_G$}\\
A_{\bar m}, \psi_{\a\bar m} &&\mbox{$(0,1)$-forms on $S$, in $\ad_G$}\\
\vf_{mn}, \chi_{\a mn} &&\mbox{$(2,0)$-forms on $S$, in $\ad_G$}
\ee
and their complex conjugates.  For the moment, take $S$ to be $\C^2$ with coordinates $z_1,z_2$.
The field $\vf$ represents transverse directions to the brane, so rotating away $S'$ corresponds to giving
$\vf$ a linear vev proportional to $T$:
\be
\vf &=& m^2 z_1 T dz_1\we dz_2
\ee

The mass scale $m^2$ is related to a characteristic (stringy) scale $M_*$ of the F-theory
compactification, and the angle $\theta$ between the branes $S,S'$.  The line $z_1=0$ is the matter curve
$\Sigma =S\cap S'$.  By supersymmetry, to find the matter on $\Sigma$, it suffices to look just for
fermionic degrees of freedom localized near $z_1=0$.

The action for fermions in the parent theory is
\be
\label{fermionaction}
I_S &=& \int_{\R^{3,1}\x S} d^4x\, \Tr\left(\chi^\a\we \bar\ptl_A \psi_\a+2i\sqrt 2 \w\we \ptl_A \eta^\a \we
\psi_\a
\right.\nonumber\\
&&\left.\qquad\qquad\qquad\quad+\half \psi^\a [\vf,\psi_\a]+\sqrt 2 \eta^\a[\bar\vf,\chi_\a]\right)
+\,\mathrm{h.c.}+\mbox{kinetic terms}
\ee
where $\w=\frac i 2 g_{i\bar \jmath}dz^i\we d\bar z^{\bar \jmath}$ is the Kahler form on $S$, and $\ptl_A =
dz^m\p{\pdr{}{z^m}+A_m}$.
 Varying with respect to $\eta$ and $\psi$ gives the zero mode equations
\be
\w\we \ptl_A \psi^\a + \frac i 2 [\bar \vf, \chi^\a] &=& 0,\\
\bar\ptl_A \chi^\a - [\vf,\psi^\a] &=& 0
\ee

The modes that get trapped near $z_1=0$ are linear combinations of
$\psi,\chi$, and $\eta$ and have a nonzero charge under the adjoint
action of $T$.   The solutions are easy to find in the absence of
flux ($A=0$, in an appropriate gauge).   They are \be \psi_{\bar
2}=0,\qquad\psi_{\bar 1},\chi_{12} &\propto& \a(z_2)e^{-|mz_1|^2}
\ee where $\a(z_2)$ is holomorphic.

The presence of flux deforms the wave functions to
\be
\psi_{\bar 1},\chi_{12} &\propto& \a(z_2)e^{-|mz_1|^2}\exp\cM(z,\bar z)
\ee
where $\cM(z,\bar z)$ is a (not necessarily holomorphic) function that vanishes when the flux vanishes.  For
instance in a constant background flux,
 $A=-F_{1\bar 1} \bar z_1 dz_1+F_{2\bar 2} z_2 d\bar z_2$,
 \footnote{Note for instance that positive flux $F_{2\bar 2}$ through the curve $z_1=0$ causes wavefunctions
to decay rapidly away from $z_2=0$.
  So, wavefunctions on $\Sigma$ are ``attracted" to regions of positive flux, which is related to the well-
known fact that the number of
   normalizable zero modes on $\Sigma$ is equal to the total flux through $\Sigma$.}
\be
\label{landaudistortion}
\cM(z,\bar z) &=& -F_{2\bar 2} z_2 \bar z_2+\half F_{1\bar 1} z_1\bar z_1+\dots
\ee

\section{Review of Yukawa Calculation}

The superpotential of the parent theory is \be
\label{superpotential} W &=& M_*^4\int_S\Tr(\bF^{(0,2)}\we \bm\Phi)
\ \ =\ \ M_*^4\int_S\Tr(\bA\we \bA\we \bm\Phi)+\mbox{quadratic} \ee
where $\bm\Phi=\vf+\theta\chi+\dots, \bA=A+\theta\psi+\dots$.  Since the
zero modes are linear combinations of variations of the superfields
$\bA$ and $\bm{\Phi}$, the cubic term in (\ref{superpotential})
gives rise to Yukawa couplings proportional to structure constants
in the parent gauge group and the wavefunction overlap of matter
fields.

\begin{figure}[!ht]
\centering
\includegraphics[width=.4\textwidth]{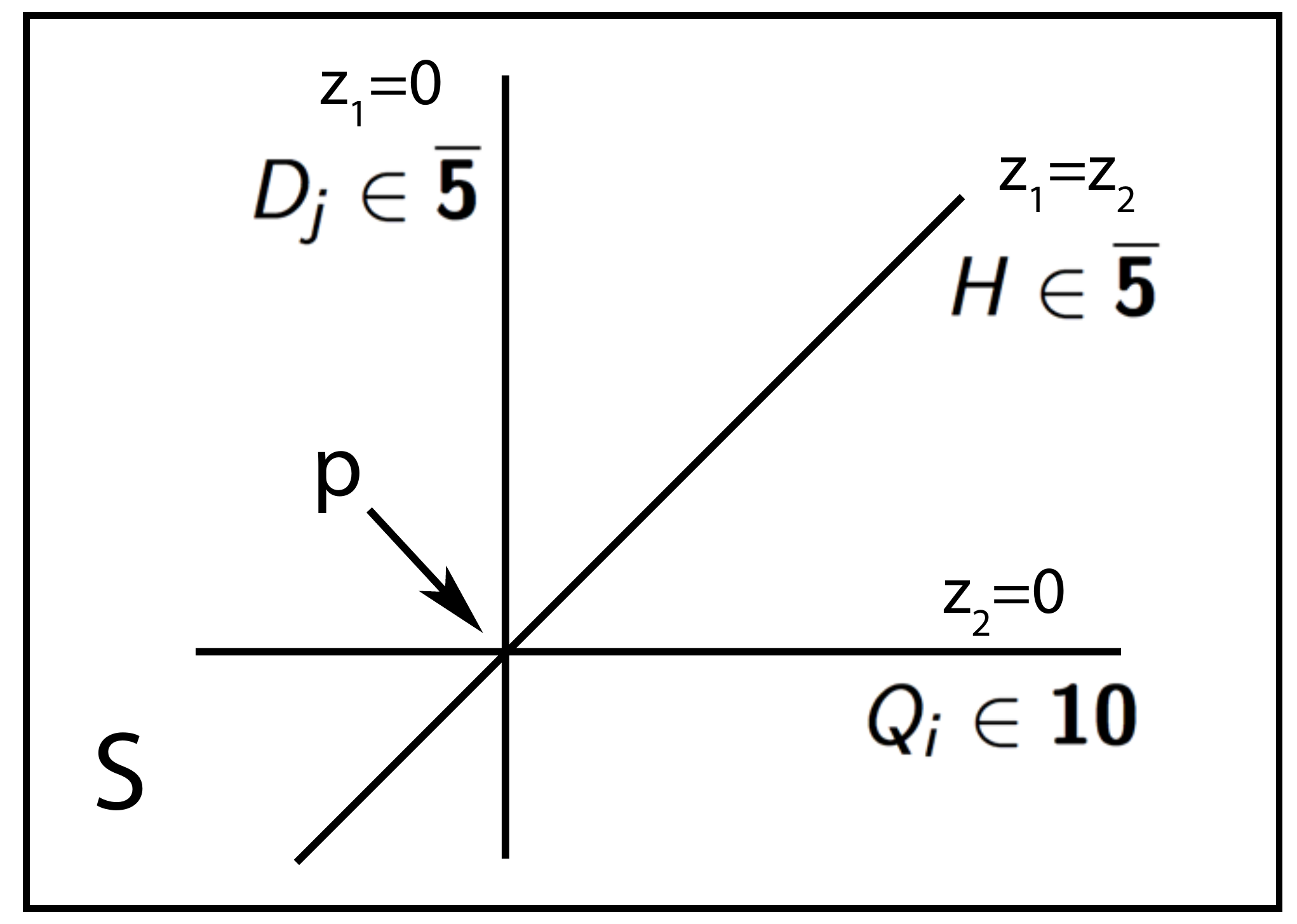}
\caption{An intersection of matter curves giving the down Yukawa.}
\label{tripleintersection}
\end{figure}

For example, consider the down Yukawa in an $SU(5)$ GUT coming from
an intersection of the matter curves
$\Sigma_Q,\Sigma_D,\Sigma_{H_d}\subset S$ at a point $p_d$.  We can
choose coordinates $z_1,z_2$ near $p_d$ such that $\Sigma_Q$ and
$\Sigma_D,$ are the zero loci of $z_2,z_1$, respectively.  For
concreteness, we'll take the curve $\Sigma_{H_d}$ to be $z_1=z_2$
(Fig. \ref{tripleintersection}).

Then the zero modes have wavefunctions
\be
\label{zeromodewavefunctions1}
Q_i &\sim& \a_i(z_1)e^{-|m_1z_2|^2+\cM_1(z,\bar z)}\\
D_j &\sim& \b_j(z_2)e^{-|m_2z_1|^2+\cM_2(z,\bar z)}\\
\label{zeromodewavefunctions3}
H_d &\sim& \g(z_1+z_2)e^{-|m_3(z_1-z_2)|^2+\cM_3(z,\bar z)}
\ee
where the $\a_i,\b_j,$ and $\g$ are holomorphic.  Following \cite{Heckman:2008qa},
by performing unitary flavor rotations we can require
\be
\a_i(z_1) &=& \p{\frac {z_1} {R_1}}^{3-i}+\mbox{lower order}\\
\b_j(z_2) &=& \p{\frac {z_2} {R_2}}^{3-j}+\mbox{lower order}\\
\g(z) &=& \mathrm{const.} + \mbox{lower order}
\ee
where $R_1,R_2$ are roughly the sizes of the matter curves $\Sigma_Q, \Sigma_D$.
 Our Yukawa is proportional to the wavefunction overlap
\be
\label{yukoverlap}
Y_{ij} &\propto& M_*^4\int d^2z_1 d^2z_2 \p{\frac {z_1} {R_1}}^{3-i}\p{\frac {z_2} {R_2}}^{3-j} \exp\p{-|mz|
^2+\cM(z,\bar z)}
\ee
where $|mz|^2$ is short for $|m_1z_2|^2+|m_2z_1|^2+|m_3(z_1-z_2)|^2$, and $\cM(z,\bar z)$
is the sum of the flux-dependent distortions $\cM_i$ for each curve.

Since the Gaussian measure $d^2 z_1 d^2 z_2 e^{-|mz|^2}$ is
invariant under $U(1)$ rotations of each coordinate, to get a
nonzero result we need to pull down sufficient powers of $\bar z$ in
the Taylor expansion \be \label{Mtaylorexpansion}
\cM(z,\bar z) &=& \sum \cM^{a,b} \p{\frac{\bar z_1}{R_1}}^a \p{\frac{\bar z_2}{R_2}}^b
+\mbox{holomorphic and mixed terms} \ee
We'll approximate $\cM^{a,b}
\sim \cM_0$, where $\cM_0$ is the
characteristic size of the
distortion $\cM$.  This is equivalent to assuming that the
distortion factor varies by order $\cM_0$ over curves of sizes $R_1$
and $R_2$.  The authors of \cite{Heckman:2008qa} identify two types
of expansions of the exponential $e^\cM$, which are important in
different limits.  The ``derivative" expansion brings down the
single term $\cM^{3-i,3-j}\bar z_1^{3-i} \bar z_2^{3-j}$ in
(\ref{Mtaylorexpansion}) necessary to make the integrand
$U(1)$-invariant, giving \be Y_{ij}^{\mathrm{DER}} &\propto&
\frac{M_*^4}{m^4}\p{\frac{1}{m^2R_1^2}}^{3-i}\p{\frac 1
{m^2R_2^2}}^{3-j} \cM_0 \ee The ``flux" expansion brings down
multiple powers of $\cM^{1,0} \bar z_1$ and $\cM^{0,1} \bar z_2$,
giving \be Y^{\mathrm{FLX}}_{ij} &\propto& \frac {M_*^4}
{m^4}\p{\frac{\cM_0}{m^2R_1^2}}^{3-i}\p{\frac {\cM_0}
{m^2R_2^2}}^{3-j} \ee

Assuming that $m\sim M_*$ and $R_1\sim R_2\sim R$, we see
that the small parameters relevant for the hierarchy in the Yukawas are
\be
\ka \ \ =\ \  \frac{1}{m^2 R^2},\qquad
\e^2 \ \ =\ \  \cM_0\ka
\ee
and we have
\be
\label{yukawahierarchies}
Y = Y^{\mathrm{FLX}}+Y^{\mathrm{DER}}
&=&
\left(
\begin{array}{ccc}
\e^8 & \e^6 & \e^4\\
\e^6 & \e^4 & \e^2\\
\e^4 & \e^2 & 1
\end{array}
\right)
+
\frac{\e^2}{\ka}\left(
\begin{array}{ccc}
\ka^4 & \ka^3 & \ka^2\\
\ka^3 & \ka^2 & \ka\\
\ka^2 & \ka   & 1
\end{array}
\right)
\ee

The relation $(RM_*)^4=\a_\GUT\inv$ implies
\be
\e \sim \ka \sim \a_\GUT^{1/2},
\ee
from which the authors of \cite{Heckman:2008qa} show that the matrices in (\ref{yukawahierarchies}) can reproduce the known quark
masses.\footnote{The success of this
estimate relies on the assumption $m\sim M_*$.  This is equivalent
to the assumption that oscillations of matter fields transverse to a
matter curve decouple at a stringy scale, which is necessary for us
to think about matter as being localized on curves in the first
place. However, the order one ratio $m/M_*$ could affect the
hierarchy in \ref{yukawahierarchies}.

Since there are order one parameters in each entry, this is not really an issue.  But the best fit from, for instance, the authors of \cite{Heckman:2008qa} for up quark masses using the $Y^\mathrm{FLX}$ hierarchy $1:\e^4:\e^8$ requires $\e\sim 0.26$, which is a bit larger than $\a_\GUT^{1/2}\sim 0.2$.  Fitting quark masses at the GUT scale requires an even bigger $\e$.  Below, when we discuss neutrino masses in terms of $\a_\GUT$, we will only be able to estimate up to similar order one factors.} Further, assuming that both
the up and down Yukawas take the form (\ref{yukawahierarchies}) with
the same basis for the left-handed quarks, they show that the
resulting quark mixing angles agree nicely with $V^\mathrm{CKM}$ in
the Standard Model.  We'll return to this assumption
in Section \ref{nearbypointssection}.

\section{Effective Field Theory Interpretation}

Notice that
the matrix $Y^{\mathrm{FLX}}$ takes the typical single-field Froggatt-Nielsen (FN)
form \cite{Froggatt:1978nt} with $U(1)$ flavor charges $4,2,0$ for generations $1,2,3$, respectively,
 and a spurion field $\e$ with flavor charge $-1$.  However, $Y^{\mathrm{DER}}$ differs
 from a usual FN structure by the small parameter $\frac {\e^2}{\ka}$ out front.
 If we wanted to reproduce only the DER matrix, we could account for it in an FN scenario
 by assigning the Higgs a charge. However, in that case the large entry $Y_{33}^\mathrm{FLX}$
 would be forbidden. Clearly, without additional fields (other than the zero-mode Higgs and three
 light generations of left- and right-handed fields), one cannot generate the sum of the two
 types of matrices with symmetries alone. One can take simplified cases where we can simply
 generate both matrices (see below), but our goal here is to write down the minimal theory
 that directly describes how the geometry produces (\ref{yukawahierarchies}).

Perhaps the simplest way to find the field theory that reproduces
the F-theory result is to explicitly do perturbation theory in the
flux from the beginning. Separate the background gauge field $A$ as
$A=A_0+a$, where $A_0$ is in the same topological class as $A$, but
has zero flux near $p_d$. In the $A_0$ background, the zero modes
have the simple holomorphic $\x$ gaussian form
(\ref{zeromodewavefunctions1}-\ref{zeromodewavefunctions3}) near
$p_d$, without the distortion factors $\cM$.  The resulting Yukawa
is that of (\ref{yukawahierarchies}) with $\e=\ka=0$, namely a
rank-1 matrix involving only the third generations.

The perturbation $a$ restores the flux near $p_d$, and somehow leads
to mixing between generations, and thus corrections to this rank-1
Yukawa. The action (\ref{fermionaction}) involves only mass mixing
so the obvious interpretation of kinetic mixing among generations is
ruled out (though it would be permitted with high-dimension terms
included in the Lagrangian). But mass mixing among the generations
is of course not permitted in a chiral theory without insertion of a
Higgs field. So the only possible interpretation is in terms of mass
mixing between zero modes and KK modes with the same gauge quantum
numbers.   (Note that we use the term KK mode for all the heavy
modes, including the orthogonal combinations of $\eta$, $\chi$, and
$\psi$.)  So we can think of $a$ as distorting zero mode
wavefunctions, or equivalently mixing zero modes of the $A_0$
background with KK modes of the $A_0$ background.

However, this still leaves the question of where the corrections to
the rank-1 Yukawa arise: if only mass mixing played a role, all the
correction terms would involve at least two powers -- rather than a
single power -- of the flux $\cM_0=\frac{\e^2}{\ka}$.  We now show
that the KK modes have nonzero Yukawa couplings to the different
generations, and these Yukawas in combination with the mixing via
$a$ are the source of the perturbation to the rank-1 Yukawa.

\subsection{Feynman Rules for KK Modes}

To make this picture precise, let's first concentrate on the left-handed quark
superfields $Q_i\in\bm{10}$.  The $Q_i$ are zero modes of a vector-like field
$(\mathbb Q,\mathbb Q^c)\in \bm{10}\oplus \bar{\bm{10}}\subset\ad_G$ on $\R^{1,3}\x S$, which has KK expansion
\be
\mathbb Q(x,z,\bar z) &=& Q_i(x)f_i(z,\bar z)+\sum_I Q_I(x) g_I(z,\bar z)\\
\mathbb Q^c(x,z,\bar z) &=& \sum_J Q^c_J(x) h_J(z,\bar z)
\ee
where the capital subscripts represent massive KK-modes.
In the absence of flux near $p_d$, the zero mode wavefunctions $f_i$ take the form $1,z_1,z_1^2$.
For the sake of computing Yukawas, we'd like to classify the KK wavefunctions $g_I$ in a similar way.
Choose a basis $g_{a,b}^{c,d}$ such that
\be
g_{a,b}^{c,d}(z,\bar z) &\sim& z_1^a z_2^b \bar z_1^c \bar z_2^d
\qquad
\mbox{near $p_d$.}
\ee
We say that the fields $Q_{a,b}^{c,d}$ associated with $g_{a,b}^{c,d}$ have ``$z$-charge $(a,b)$" and
``$\bar z$-charge $(c,d)$."  The zero modes are a special case: $Q_i=Q_{i,0}^{0,0}$.    For the KK modes
 with $z$-charge $(0,0)$,  we'll use a simpler notation $Q^{c,d}=Q^{c,d}_{0,0}$.

Yukawa couplings are nonzero only when the powers of $z_i$ are the same as
the powers of $\bar z_i$.  And each insertion of $z_i\bar z_i$ in the integral
gives a factor $\ka_i=\frac {1}{m^2 R_i^2}$.  Thus, the allowed Yukawas are
\be
\label{allowedyukawas}
H_{a_1,b_1}^{c_1,d_1}Q_{a_2,b_2}^{c_2,d_2} D_{a_3,b_3}^{c_3,d_3}
\ka_1^{a_1+a_2+a_3}\ka_2^{b_1+b_2+b_3},
\qquad\mbox{where $\sum a_i=\sum c_i$ and $\sum b_i=\sum d_i$}
\ee
For instance, at leading order in  powers of $z$ and $\bar{z}$, the Yukawas
involving two zero modes and one KK mode are of the form

\be
\begin{array}{ccc}
\begin{fmffile}{HKKYuk}
\begin{fmfgraph*}(40,40)
\fmfset{arrow_len}{3mm}
\fmfleft{i1}
\fmfright{o1,o2}
\fmf{double}{i1,v1}
\fmf{plain}{o1,v1,o2}
\fmfdot{v1}
\fmflabel{$H^{a,b}$}{i1}
\fmflabel{$Q_{a,0}$}{o1}
\fmflabel{$D_{0,b}$}{o2}
\fmflabel{$\ka_1^a \ka_2^b$}{v1}
\end{fmfgraph*}
\end{fmffile}
\qquad&\qquad\qquad
\begin{fmffile}{QKKYuk}
\begin{fmfgraph*}(40,40)
\fmfset{arrow_len}{3mm}
\fmfleft{i1}
\fmfright{o1,o2}
\fmf{plain}{i1,v1}
\fmf{plain}{o2,v1}
\fmf{double}{v1,o1}
\fmfdot{v1}
\fmflabel{$H$}{i1}
\fmflabel{$Q^{0,b}$}{o1}
\fmflabel{$D_{0,b}$}{o2}
\fmflabel{$\ka_2^b$}{v1}
\end{fmfgraph*}
\end{fmffile}
\qquad\qquad&\qquad
\begin{fmffile}{DKKYuk}
\begin{fmfgraph*}(40,40)
\fmfset{arrow_len}{3mm}
\fmfleft{i1}
\fmfright{o1,o2}
\fmf{plain}{i1,v1}
\fmf{double}{o2,v1}
\fmf{plain}{v1,o1}
\fmfdot{v1}
\fmflabel{$H$}{i1}
\fmflabel{$Q_{a,0}$}{o1}
\fmflabel{$D^{a,0}$}{o2}
\fmflabel{$\ka_1^a$}{v1}
\end{fmfgraph*}
\end{fmffile}
\end{array}
\ee
\vspace{.1in}

The shapes of the wavefunctions enforce the relations between $z$
and $\bar z$ charges and the number of powers of $\ka_i$.  We can
mimic this with flavor charges by declaring that $\ka_i$ have
charges $-1,-1$ under $z_i,\bar z_i$, respectively.  Then the
couplings (\ref{allowedyukawas}) are enforced by charge invariance.
 In other words, we can use $U(1)$ symmetries as a trick to get the right structure, but it really comes from
the geometry.

Now let's combine this with the mixing from the perturbation $a$.
We can read off the KK mass matrix from the terms bilinear in the fermions
$q_I,q^c_J$ associated with $Q_I$ and $Q_J^c$.  These are linear combinations of
the fields $\psi,\chi,\eta$ in the parent theory, so $M^\KK$ comes from fermion bilinear
terms in the action (\ref{fermionaction}) with gauge field $A_0$:
\be
I_{A_0} &=& \int_{\R^{3,1}\x S} d^4x\, \Tr\left(\phantom\half\!\!\!\chi^\a\we
\bar\ptl_{A_0} \psi_\a+2i\sqrt 2 \w\we \ptl_{A_0} \eta^\a \we \psi_\a\right.
\nonumber\\
&& \qquad\qquad\quad
\qquad
\left.+\half \psi^\a [\vf,\psi_\a]+\sqrt 2 \eta^\a[\bar\vf,\chi_\a]\right)+\,\mathrm{h.c.}\\
&=& \int d^4 x\, M^{(\KK)}_{JI} q^c_J q + \mathrm{h.c.}
\ee
Similarly, the gauge field perturbation $a$ couples to fermions as follows
\be
I_a &=& \int_{\R^{3,1}\x S} d^4x\, \Tr\p{\chi^\a\we \bar a\we  \psi_\a+2i\sqrt 2 \w\we a \we \eta^\a \psi_\a}+
\,\mathrm{h.c.}\\
&=& \int d^4x\, a_{JI}q_J^c q_I + a_{Ji}q_J^c q_i+ \mathrm{h.c.}
\ee
Note that while $M^{(\KK)}$ clearly only couples to KK modes, the perturbation $a$
couples KK modes to each other, and also KK modes of $\mathbb{Q}^c$ to zero modes of $\mathbb{Q}$.
 Including this mixing and integrating out the $Q_J^c$ to lowest order in momentum induces a Yukawa coupling
between zero modes alone
\be
a_{Ji}
\contraction{}{Q}{{}^c_J Q_i \quad Y_{Ij} H}{Q}
Q^c_J Q_i \quad Y_{Ij} H Q_I D_j
&=&
(M^{\KK})^{-1}_{IJ}a_{Ji} Y_{Ij} H Q_i D_j
\ee
The matrix
\be
(M^{\KK})^{-1}_{IJ}a_{Ji}
\ee
is just multiplication by the distortion factor $\cM(z,\bar z)$ that comes from turning on $a$.
This may be a bit surprising, since our notation has obscured the $z$ dependence of all the wavefunctions,
but it must be true since multiplication by $(M^\KK)^{-1} a$ and $\cM(z,\bar z)$ both give the first order
variation in the zero modes from turning on $a$.  For instance, in our previous example
(\ref{landaudistortion})
 with matter curve $z_1=0$ in $\C^2$, if we start with $A_0=0$, and turn on a constant flux $a=F_{2\bar 2} z_2
d\bar z_2$, then
\be
(M^{\KK})^{-1}_{IJ}a_{Ji}
&\sim& \bar\ptl_2\inv a = F_{2\bar 2} z_2 \bar z_2
\ee
which is indeed $\cM$ associated with constant flux $F_{2\bar 2}$ through the curve.
In the Taylor expansion (\ref{Mtaylorexpansion}), the derivatives $\cM^{a,b}$ allow the
zero modes to mix with KK modes that carry $\bar z$-charge $(a,b)$ (in effect compensating
for the apparent violation of $\bar{z}$ charge that allows the Yukawas to be nonzero).
 Thus, we have the following diagrammatic rule for mixing to KK modes and propagating:
\be
\parbox{45mm}{\begin{fmffile}{zerotoKK}
\begin{fmfgraph*}(120,40)
\fmfset{arrow_len}{3mm}
\fmfleft{i1}
\fmfright{o1}
\fmf{plain}{i1,v1}
\fmf{double,label=$\Labc$}{v1,o1}
\fmfdot{v1}
\fmflabel{$Q_a$}{i1}
\end{fmfgraph*}
\end{fmffile}}
&=& \cM^{b,c}
\\
\parbox{45mm}{\begin{fmffile}{KKtoKK}
\begin{fmfgraph*}(120,40)
\fmfset{arrow_len}{3mm}
\fmfleft{i1}
\fmfright{o1}
\fmf{double}{i1,v1}
\fmf{double,label=$\Labcdef$}{v1,o1}
\fmfdot{v1}
\fmflabel{$Q_{a,b}^{c,d}$}{i1}
\end{fmfgraph*}
\end{fmffile}}
&=& \cM^{e,f}
\ee
where the factors $\cM^{b,c}$ include both the vertex and the $KK$ propagator,
since those elements will appear together in all our Feynman diagrams.

\subsection{Diagrammatic Interpretation of $Y^\mathrm{DER}$ and $Y^\mathrm{FLX}$}

The two Yukawa textures $Y^{\mathrm{DER}}$ and $Y^{\mathrm{FLX}}$ arise naturally
 from the charge assignments, and have a diagrammatic interpretation in terms of mixing through KK modes

\begin{eqnarray*}
\mathrm{DER}: && H Q_a D_b \cM^{a,b}\ka_1^a \ka_2^b\\
&&
\parbox{40mm}{\begin{fmffile}{DER}
\begin{fmfgraph*}(170,80)
\fmfset{arrow_len}{3mm}
\fmfleft{i1}
\fmfright{o1,o2}
\fmf{plain}{i1,v1}
\fmf{double,label=$\Laab$}{v1,v2}
\fmfdot{v1}
\fmflabel{$Q_a$}{i1}
\fmf{plain}{o1,v2,o2}
\fmfdot{v2}
\fmflabel{$D_b$}{o1}
\fmflabel{$H$}{o2}
\fmflabel{$\ka_1^a \ka_2^b$}{v2}
\end{fmfgraph*}
\end{fmffile}}
\hspace{2in}
\\
\\
\\
\mathrm{FLX}: && H Q_a D_b (\cM^{1,0}\ka_1)^a(\cM^{0,1} \ka_2)^b\\
\\
\\
&&
\parbox{60mm}{\begin{fmffile}{FLX}
\begin{fmfgraph*}(250,150)
\fmfset{arrow_len}{3mm}
\fmfleft{i1}
\fmfright{o1}
\fmftop{o2}
\fmf{plain}{i1,v1}
\fmf{double,label=$\Lflxa$}{v1,v2}
\fmf{double,label=$\Lflxb$,l.side=left}{v2,v3}
\fmflabel{$Q_a$}{i1}
\fmf{plain}{v3,o2}
\fmf{plain}{o1,v4}
\fmf{double,label=$\Rflxa$,l.side=right}{v4,v5}
\fmf{double,label=$\Rflxb$,l.side=right}{v5,v3}
\fmflabel{$D_b$}{o1}
\fmflabel{$H$}{o2}
\fmfv{l=$\ka_1^a\ka_2^b$,l.a=-90,l.d=.03w}{v3}
\fmfdot{v1,v2,v3,v4,v5}
\end{fmfgraph*}
\end{fmffile}}
\end{eqnarray*}
\vspace{-1in}

Note that in both cases, the zero modes carried the $z$ charge that is also carried by
the KK mode whereas the KK-modes (that mix with the zero modes) carry the $\bar{z}$ charges
 that permit Yukawas to be nonvanishing.\footnote{The DER scenario is similar to shining in
  that a different spurion $\cM^{a,b}$ is responsible for each entry of the matrix.
  However, the hierarchy in entries comes from $z$ and $\bar{z}$ charges, rather than
   signal propagation over extra dimensions.}

Of course, we also get corrections to the Yukawas through mixing
with Higgs KK modes, in which case the KK modes carry purely $\bar{z}$ charge.
 For instance, the flux interaction can come from
\be
\parbox{125mm}{\begin{fmffile}{FLXHiggs}
\begin{fmfgraph*}(370,80)
\fmfset{arrow_len}{3mm}
\fmfleft{i1}
\fmfright{o1,o2}
\fmf{plain}{i1,v1}
\fmf{double,label=$\Hone$}{v1,v2}
\fmf{double,label=$\Htwo$}{v2,v3}
\fmf{double,label=$\Hthree$}{v3,v4}
\fmf{double,label=$\Hfour$}{v4,v5}
\fmfdot{v1,v2,v3,v4,v5}
\fmflabel{$H$}{i1}
\fmf{plain}{o1,v5,o2}
\fmfdot{v2}
\fmflabel{$D_b$}{o1}
\fmflabel{$Q_a$}{o2}
\fmflabel{$\ka_1^a \ka_2^b$}{v5}
\end{fmfgraph*}
\end{fmffile}}
\ee
plus permutations of the internal KK mode lines.

Notice that if the expansion parameters $\epsilon^2$ and
 $\kappa\sim\ka_1\sim\ka_2$ were the same, we could have reproduced
 the sum $Y^\mathrm{FLX}+Y^\mathrm{DER}$ by introducing a single charged
 Higgs field, which could couple in the DER matrix while a neutral Higgs field coupled
 in the FLX matrix. However, ref. \cite{Heckman:2008qa} matched quark masses with $\e,\ka$ as
  independent parameters. Furthermore, our goal here was to answer quite generally
  the question of effective field theory underlies the F-theory form found in \cite{Heckman:2008qa}.

We also note that in principle we could reproduce the above form
with only
 a single generation of KK modes -- either flavored ones for the Higgses or a single
 additional copy for all of the fermions. One can have a model with just
 such copies that reproduces the F-theory form, without the full KK structure and the full extra-dimensional
theory.

\subsection{Froggatt-Nielsen Type Models}

We can account for the Yukawas purely in terms of charges and spurions
(without additional Higgs or fermion fields) by integrating out the KK modes from our model.
The result isn't pretty.  We have four $U(1)$ charges $(a,b,c,d)$, counting the powers of $z_1,z_2,\bar
z_1,\bar z_2$,
respectively.  We should think of $\cM^{a,b}$ as spurions with charges $(0,0,a,b)$, and $\ka_1$ and $\ka_2$ as
 spurions with charges $(-1,0,-1,0)$ and $(0,-1,0,-1)$, respectively.  The zero modes $Q_a$ have charges
 $(a,0,0,0)$ and the $D_b$ have charges $(0,b,0,0)$.

In the situation where $\ka_1\aeq \ka_2$, we can simplify the charges and spurions
 while sacrificing some fidelity to the F-theory geometry.  Specifically, we no longer
 need to distinguish between $z_1$ and $z_2$, and we can just keep track of total $z$-charge and
 total $\bar z$-charge.  We need only one spurion $\ka$ with charges $(-1,-1)$, and four
 spurions $\cM^a$ with charges $(0,a)$ for $a=1,2,3,4$.  Both the right- and left-handed
 generations have charges $(i,0)$ for $i=0,1,2$.  This is sufficient to reproduce both
  Yukawa textures of (\ref{yukawahierarchies}).

\section{Mixing Matrices From Nearby Intersection Points}
\label{nearbypointssection}

So far we have included only the perturbative contributions to the Yukawa couplings
arising from flux-induced distortions of the wavefunctions.  If the up and down Yukawas
are generated at the same point, the resulting mixing matrix $U_\mathrm{same}$ agrees
well with $V^\mathrm{CKM}$ in the Standard Model \cite{Heckman:2008qa}.  But in general
the intersections occur at separated points $p_u$ and $p_d$.  Then, even though each individual
 Yukawa matrix is rank-one, the associated eigenvectors will in general be misaligned,
 yielding an additional contribution  $U_\mathrm{sep}$ to the mixing matrix. Since we
  already know $V^\mathrm{CKM}$ agrees well with $U_\mathrm{same}$, this yields a rough
  constraint on the separation $|p_u-p_d|$, which we now derive.

Around any point $p\in\Sigma_q$, we can pick coordinates $z$ such
that $z=0$ at $p$ and choose an orthonormal basis of zero modes
$f_p^i$ such that \be f^i_p(z)&\sim& z^{3-i}+\mbox{lower order}. \ee
The basis $f_p^i$ is unique (up to $U(1)$ rotations of each element)
at each point $p$. The matrix $U_\mathrm{sep}$ is the unitary
rotation between the zero mode bases $f^i_{p_u}$ and $f^j_{p_d}$. If
$p_u$ and $p_d$ are far apart on the quark curve $\Sigma_Q$, these
bases will in general be unrelated and $U_\mathrm{sep}$ will have
order one angles. This is clearly undesirable in the quark sector
where mixing angles are small.

The most obvious way to avoid this large mixing
contribution is to require $p_u$ be near $p_d$. In this case, we can say something
 nontrivial about $U_\mathrm{sep}$ by keeping track of the way the basis $f_p$
 varies with small changes in $p$.  The key observation is that as we move from
 $z=0$ to a nearby point, the functions $1$ and $z$ mix with each other at first
 order, as do the functions $z$ and $z^2$.  However, $1$ and $z^2$ do not mix at first order.

By requiring that the $f^i_p$ vanish with the appropriate degree at $p$, we can
derive the lower triangular part of an evolution equation for $f_p^i$:
\be
\pdr{}{p}
\p{
\begin{array}{c}
f_p^3\\
f_p^2\\
f_p^1
\end{array}}
&=&
\left.\p{
\begin{array}{ccc}
0 & ? & ?\\
-\frac{\ptl_z f_p^2}{f_p^3} & 0 & ?\\
0 & -\frac{\ptl^2_z f_p^1}{\ptl_z f_p^2} & 0
\end{array}}\right|_{z=p}
\p{
\begin{array}{c}
f_p^3\\
f_p^2\\
f_p^1
\end{array}}
\ee
We've set the diagonal entries to zero because they would contribute just a $U(1)$
rotation on each basis element.\footnote{Note that $\ptl_z f$ is gauge-covariant at
a point where $f$ vanishes, since $\ptl_z f=(\ptl_z+A_z) f$ at that point.  Thus,
the ratio $\frac{\ptl_z f^2}{f^3}$ is well-defined. For the same reason, the second
derivative $\ptl_z^2 f^3$ is gauge-covariant, and the ratio $\frac{\ptl_z^2 f^1}{\ptl_z f^2}$ is well-defined.}
Since the basis should remain orthogonal, we can fill in the upper triangular part with the requirement that
the $3\x3$ matrix above be anti-Hermitian:\footnote{We can think of this equation as defining a $U(3)$ connection on the curve $\Sigma_q$.  Since the basis $f^i_p$ is unique up to $U(1)$ rotations of each basis element, the curvature of this connection should be contained within a $U(1)^3$ subgroup.}
\be
\label{basischange}
\pdr{}{p}
\p{
\begin{array}{c}
f_p^3\\
f_p^2\\
f_p^1
\end{array}}
&=&
\left.\p{
\begin{array}{ccc}
0 & \bar{\frac{\ptl_z f_p^2}{f_p^3}} & 0\\
-\frac{\ptl_z f_p^2}{f_p^3} & 0 & \bar{\frac{\ptl_z^2 f_p^1}{\ptl_z f_p^2}}\\
0 & -\frac{\ptl_z^2 f_p^1}{\ptl_z f_p^2} & 0
\end{array}}\right|_{z=p}
\p{
\begin{array}{c}
f_p^3\\
f_p^2\\
f_p^1
\end{array}}
\ee

Finally, estimating $\ptl_z\sim R\inv$, and integrating this over a
short distance $\de \. R$ gives the mixing matrix \be
\label{pointsepmixing} U_\mathrm{sep} &\aeq& \left(
\begin{array}{ccc}
1 & \de & \de^2\\
\de & 1 & \de\\
\de^2 & \de & 1
\end{array}
\right)
\ee
Thus, nearby points have order $\de$ mixings between adjacent generations, but order
 $\de^2$ mixings between the 1st and 3rd generations.\footnote{This form (\ref{pointsepmixing})
 is familiar from the simplest Froggatt-Neilsen flavor models \cite{Froggatt:1978nt}.}
  This is not the structure of $V^{CKM}$ in the standard model, and we'd like to ensure
  it gives a negligible contribution to mixing compared to $U_\mathrm{same}$.
  The strongest bound on $\de$ comes from the fact that the mixing between the
  second and third generations in the standard model is small $V^{CKM}_{23}\aeq 0.04$.
  We should have $\de\lesssim 0.04$, or $|p_u-p_d|\lesssim0.04 R$.\footnote{This is a
  slightly stronger constraint than the rough estimate in \cite{Heckman:2008qa}}

Our constraint supports the observation in \cite{Heckman:2008qa}
that the quark sector shows either nontrivial fine tuning or perhaps
evidence of a higher unification structure that forces $p_u$ near
$p_d$.

\section{The Neutrino Sector}

\subsection{Mixing Angles for Dirac Neutrinos}

Though dangerous with respect to quarks, the mixing matrix
(\ref{pointsepmixing}) is interesting from the perspective of Dirac
neutrinos.  Mixing angles for neutrinos follow a very different
pattern than those for quarks.  The current bounds are
\cite{Amsler:2008zzb} \be \sin^2(2\theta_{23})\ \ >\ \ 0.92 ,\qquad
\sin^2(2\theta_{12})\ \ =\ \ 0.86^{+0.03}_{-0.04} ,\qquad
\sin^2(2\theta_{13})\ \ <\ \ 0.19 \ee We first note that the angles in
(\ref{pointsepmixing}) take the basic form observed for neutrinos,
with large mixing between adjacent generations and suppressed mixing
between the first and third generations.\footnote{A single field Froggatt-Nielsen model could also give mixing angles with the texture (\ref{pointsepmixing}).}
 Of course once $p_u$ and
$p_d$ are sufficiently far away, $\de$ is not small and all angles
are of the same order.  We can safely say that with distant points
we predict large mixing angles (this point was also made in ref.
\cite{Heckman:2008qa}). We can also view (\ref{pointsepmixing}) as
suggesting that $\theta_{13}$ might be somewhat smaller if the Yukawas
for leptons and neutrinos are generated at points that are not too
far apart.   These models would nonetheless be strongly disfavored
if experiment determines that $\theta_{13}$ is substantially smaller
than $\theta_{12}^2$.  However, there are recent indications \cite{SanchezTalk} of a lower bound on $\theta_{13}$, and a best fit consistent with the relation $\theta_{13}\sim \theta_{12}^2$.

\subsection{Neutrino Masses}

We now turn to the question of the masses themselves. We  know only
neutrino mass-squared differences, but they already look very
different from those in the the quark sector \cite{Amsler:2008zzb}
\be \label{pdg1}
\De m_{21}^2 &=& (8.0\pm 0.3)\x 10^{-5} \eV^2\\
\label{pdg2} \De m_{32}^2 &=& (2.5\pm 0.5)\x 10^{-3} \eV^2 \ee In
addition to the obvious fact that neutrino masses are  much smaller
than either quark or charged lepton masses, the ratios of masses is
also quite different.

 Firstly, if neutrino masses fall into a normal
hierarchy, with for instance $m_1<m_2<m_3$, then (\ref{pdg2})
suggests that the ratio $m_3/m_2$ is of order $5$, not $25$ like for
quarks at the GUT scale.  In this case, we would need to explain why
the neutrino hierarchy is less steep.  Secondly, it's possible that
neutrino masses fall into an ``inverted hierarchy," with $m_1,m_2\gg
m_3$. This pattern  would certainly not conform to the rank-one
structure of the quark mass matrices.

Particularly in the case of an inverted hierarchy, though likely
also in the case of the normal hierarchy given the mass ratios, we
need a significantly different Yukawa structure from the F-theory
motivated structures that we have  encountered up to this point. We
now argue that for an $SU(5)$ GUT, the singlet nature of the
neutrino makes this a reasonable possibility.  In particular, we
show that whereas the quark matrix is expected to be rank-1, because
the right-handed neutrino is a singlet, the neutrino mass matrix can
generically have two large eigenvalues.

\subsubsection{A Single Intersection Point}

\begin{figure}[!ht]
\centering
\includegraphics[width=.4\textwidth]{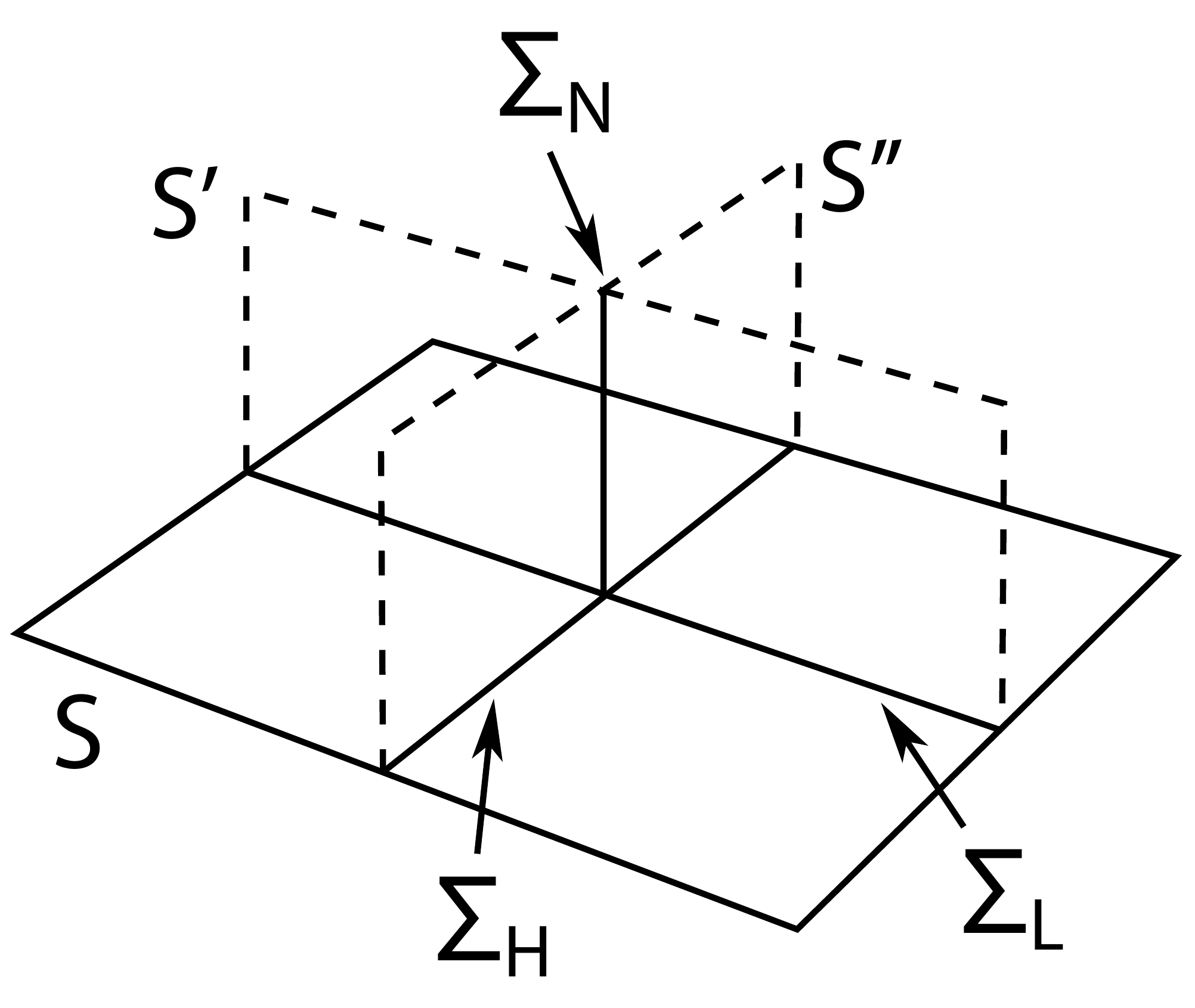}
\caption{An approximately rank-1 Yukawa involving a gauge singlet.}
\label{neutrinoyuk}
\end{figure}

The fact that right-handed neutrinos $N$ are gauge singlets means
that the matter curve
 $\Sigma_N$ must lie off the GUT brane $S$. By
contrast, the Higgs $H$ and leptons $L$ have nontrivial gauge
charges, so they live on  curves inside $S$.  Suppose $L$ and $H$
are localized on $\Sigma_L=S'\cap S$ and $\Sigma_H=S''\cap S$,
respectively.  There is now the possibility that  the singlet
neutrinos live on $\Sigma_N=S'\cap S''$. In this case, each
intersection point in $S\cap S'\cap S''$ would generate the Yukawa
$NHL$. We'll assume that this is the case and that $\Sigma_N$
supports three zero modes (Fig. \ref{neutrinoyuk}).

If $S\cap S'\cap S''$ is a single point, we get an approximately
rank-$1$ Dirac mass
matrix $m$ of the form $Y^\mathrm{FLX}$ or
$Y^\mathrm{DER}$, just like in the quark sector.  This is difficult
to reconcile with experiment, whether neutrinos are Majorana or
Dirac.  We briefly explain why, and then discuss a more viable
setup.

If neutrinos are Dirac, we could explain their small masses via
localization of right-handed neutrino wavefunctions away from the
GUT brane $S$.\footnote{Based on the curvature of the geometry near
$S$, it's possible to estimate the resulting exponential suppression
in the size of the $HNL$ coupling \cite{Beasley:2008kw}.}  But then
the neutrino mass matrix would generically have too large a
hierarchy.

Suppose instead that $N$ gets a large Majorana mass $M$, either from
a self-intersection $\Sigma_N\cap \Sigma_N \cap \Sigma_\Theta$ with
a scalar $\Theta$ that gets a vev, or (perhaps more interestingly)
via $D3$-instantons wrapping one of the branes that $\Sigma_N$ lies
in \cite{Heckman:2008es,Marsano:2008py}.\footnote{Another recently considered possibility is that right-handed Majorana masses are KK masses \cite{FutureHV}.}

In the case of a $\Sigma_N$ self-intersection at a point $p$, the
matrix $M$ will have eigenvalues of order $1,1,\e^2$.  To see why,
we first resolve the self-intersection at $p$, thinking of
$\Sigma_N$ as the image of a smooth curve $\tl \Sigma_N$ with two
points $p_1,p_2\in \tl \Sigma_N$ that both map to $p$.  As before,
we can choose a basis of wavefunctions $f^{i}_{p_1}$ that look like
$1,z,z^2$ near $p_1$.  However, that uses up our freedom to redefine
our basis, and generically all the
$f^i_{p_1}$ will be constant to leading order at $p_2$. Thus, in
this basis the wavefunction overlap integral gives a symmetric
matrix of the form \be M&\sim& \p{
\begin{array}{ccc}
\e^4 & \e^2 & 1\\
\e^2 & \e^2 & 1\\
1 & 1 & 1
\end{array}
}+\frac{\e^2}{\ka}\p{
\begin{array}{ccc}
\ka^2 & \ka & 1\\
\ka & \ka & 1\\
1 & 1 & 1
\end{array}}
\ee If $\e$ were zero, $M$ would be rank-$2$.  Since $\det(M)\sim
\e^2$, the
third eigenvalue is of order $\e^2$.

An alternative source of Majorana masses is instanton effects.  $M$
in this case is expected to be anarchic, with all eigenvalues of
order unity \cite{Heckman:2008es}.

Either way, the Majorana contribution might be relevant to the
overall size of the neutrino masses, but does not help with the
hierarchy. Only if Majorana masses aligned with Dirac masses to
cancel the hierararchies in the Dirac masses could they help make
the neutrino masses fit observations better. Instead, for the
Majorana masses described above,  the neutrino mass matrix $m^T
M^{-1} m$ will inherit too large a large hierarchy from the Dirac
mass $m$, in contradiction with experiment.

\subsubsection{Multiple Intersection Points}

If $S',S''$ were straight and completely orthogonal to $S$, there
would be a single intersection point in $S\cap S'\cap S''$.  However
there could be more, and each would contribute to the Yukawa
coupling $NHL$.  From the point of view of the matter curves, having
multiple triple-intersection points looks extremely nongeneric.
However, since the curves arise from pairwise intersections of
surfaces, they can't be moved around independently and this
situation is perfectly natural and stable under perturbations.

\begin{figure}[!ht]
\centering
\includegraphics[width=.5\textwidth]{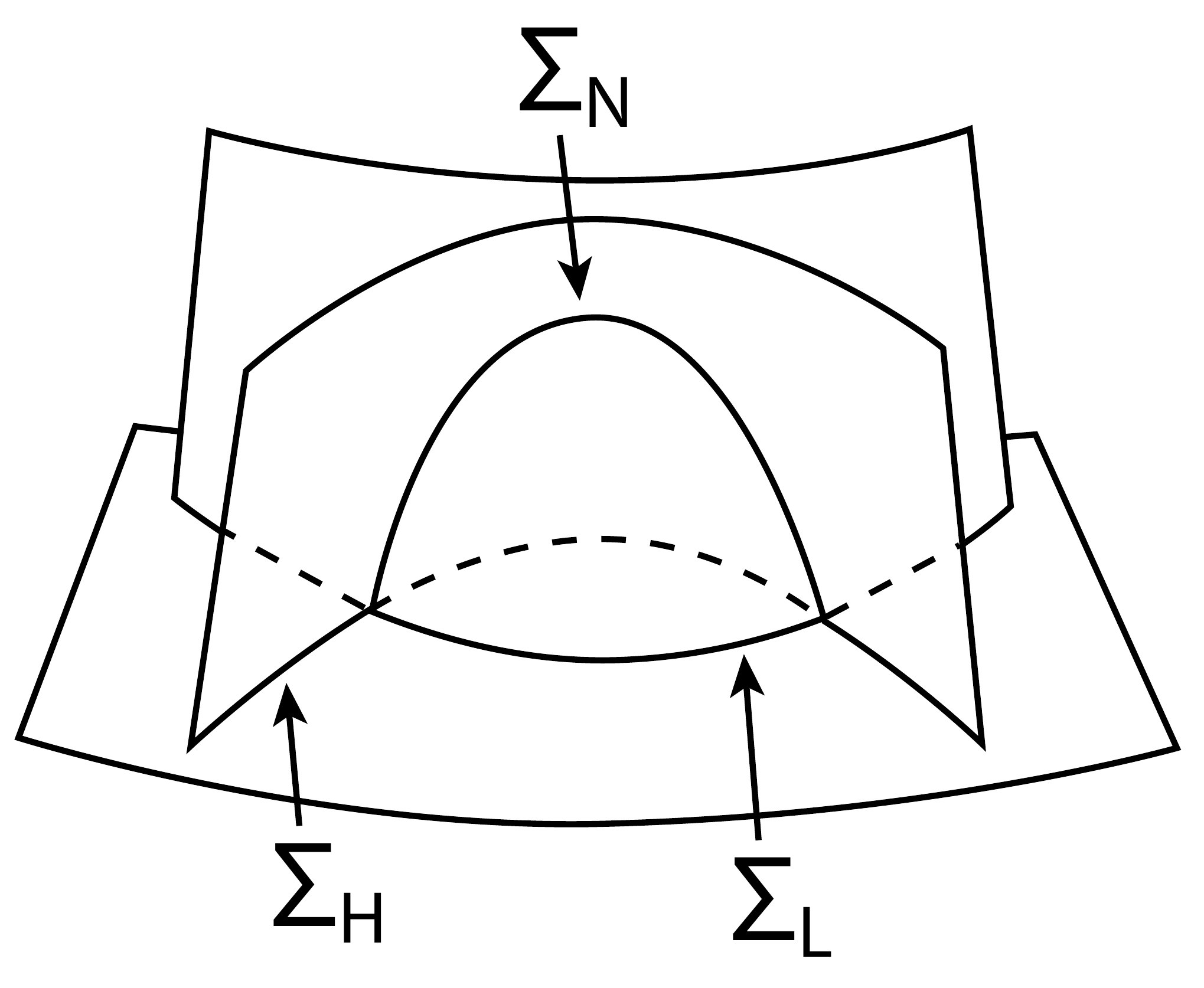}
\caption{An approximately rank-2 Yukawa involving a gauge singlet.}
\label{neutrinoyuk2}
\end{figure}

This contrasts with the quark sector, where a triple intersection of
curves $\Sigma_{H_d}\cap \Sigma_Q\cap \Sigma_D$ inside $S$ is
already nongeneric, relying on the assumption of an enhancement to a
higher rank gauge  symmetry at the intersection point.  There's no
reason to expect that the curves $\Sigma_{H_d},\Sigma_Q,\Sigma_D$
might join again at a different symmetry enhancement point.  So the
prediction of a rank-one Yukawa should be robust for the quarks.
However, the fact that neutrinos are gauge singlets implies that the
geometry giving rise to their interactions is qualitatively
different.

Interpreting the Yukawas in terms of the enhanced gauge symmetry, we
are saying the quark Yukawas require an enhancement to an
exceptional group, whereas the lepton Yukawa does not. Although locally one might generate an
exceptional group, it is clearly less generic for the global
geometry to allow two enhanced symmetry points. There is no such
nongeneric situation required for the neutrino mass.

The Dirac mass resulting from multiple intersections of $S,S',S''$
is then a sum of (approximately) rank-1 matrices $m_q$ for each
point $q\in S\cap S'\cap S''$.  With two intersection points,
perhaps the most likely of the possibilities, we generically get a
rank-2 matrix. This can accommodate either a normal or inverted
hierarchy. For instance, if $q_1$ is near $q_2$ along the curves
$\Sigma_{L},\Sigma_H,\Sigma_N$, then $m_{q_1}$ and $m_{q_2}$ will be
approximately aligned, since the appropriate choice of orthonormal
bases $f^i$ on each curve is similar at $q_1$ and at $q_2$ (as in
Section \ref{nearbypointssection}).  The degree of alignment depends
on the distance $|q_1-q_2|/R=\de$ along the matter curves.
Generically, we expect $m_{q_1}+m_{q_2}$ to have eigenvalues of
order $1,\de^2,\e^2$.  That is, the second eigenvalue is set by the
separation $\de$, and the third eigenvalue is set by the flux
$\e^2$, and vanishes when the flux goes to zero.  Taking $\de \aeq
1/\sqrt 5$ produces a viable normal hierarchy. Notice that $\e^2$ is
smaller than the quark mass hierarchy because although the natural basis for neutrino wavefunctions
might be of the form $1$, $z$, $z^2$ near the first Yukawa, there is
no reason for this to be true (given our choice of $z$) at the
second Yukawa.  Whereas in the quark case, both left- and right-handed wavefunctions
simultaneously lead to suppression, in this case, there is only one
source of suppression.

So if $q_1$ and $q_2$ are not close, so that $m_{q_1}$ and $m_{q_2}$
are
generically misaligned, then we expect two order one eigenvalues,
and a third of order $\e^2$.  When the large eigenvalues are very
nearly degenerate, this could reproduce the mass spectrum required
for an inverted hierarchy. In a slightly nongeneric case, the two
large eigenvalues might differ by a factor of five and produce a
normal hierarchy with the lowest eigenvalue much smaller than the
other two.

In either of these cases, if $N$ gets an anarchic Majorana mass from
instanton effects, the
resulting neutrino masses will still be
approximately rank-2, with one small eigenvalue of order $\e^2\sim
\a_\GUT\sim \frac 1 {25}$ relative to the biggest.\footnote{This same prediction for the smallest eigenvalue was also recently derived from a very different geometrical setup for neutrinos in F-theory, though only with a normal hierarchy \cite{FutureHV}.  Our upper bounds on measurements of $m_{\b\b}$ and $m_\nu$ will apply to their scenario as well.}

 So it seems a fairly robust prediction of this scenario is a rank-2
 mass matrix for the neutrinos at leading order.
Conservatively, the third eigenvalue, either in the normal or inverted hierarchy,
should be at least a factor of ten smaller than the largest
eigenvalue. In either normal or inverted scenarios, this sets the
overall scale for neutrino masses -- not just their mass difference.

It is of interest to ask whether the approximately rank-2 form for
the neutrino mass matrix
 could be tested. The best measurement in this regard could be neutrinoless double beta decay (though possible cosmological measurements might ultimately test the overall neutrino mass scale $m_\nu=m_1+m_2+m_3$ down to $0.04\eV$ \cite{Kaplinghat:2003bh,Hannestad:2002cn}).  The matrix element for the decay is proportional to the element $m_{\b\b}$ in the neutrino mass matrix.  We have \cite{Vogel:2006sq}
\be
|m_{\b\b}|\approx |\cos^2(\theta_{12}) m_1+\e^{i \a_{12}} \sin^2(\theta_{12})m_2+e^{i\a_{13}} \sin^2
(\theta_{13}) m_3|
\ee
where $\a_{12}$ and $\a_{13}$ are possibly nonvanishing Majorana phases.

The size of $|m_{\b\b}|$ depends on whether we have a normal or
inverted hierarchy. Assuming near-vanishing $m_1$, with the normal
hierarchy, the contribution comes primarily from $m_2$. But
for
nonzero $\theta_{13}$, there can be a reasonably large correction since
the largest eigenvalue $.046\eV<m<.056\eV$ is significantly bigger then the middle eigenvalue $.0088\eV<m<.
0091\eV$.

 We then find
\be
0\eV<|m_{\b\b}|<.005\eV
\ee
where most of the uncertainty comes from the unknown phase
$\a_{12}-\a_{13}$. If $m_1$ is nonvanishing, there will be an
additive contribution. A reasonable estimate for this contribution
in our scenario is of order $\a_\GUT m_3 \approx 0.002\eV$. If, on the other
hand, $m_1 \approx m_2$, as might be more generically the case (that
is, not in our models), we would expect an additional contribution
of order $0.01\eV$ (since there is a linear contribution from $m_1$ not
suppressed by $\sin^2\theta_{12}$). Distinguishing the small $m_1$ value of our model from the generically
larger one requires a level of precision of order $0.01\eV$, which is
clearly beyond the capacity of any planned experiment. This
nonetheless provides a useful target if we are to ultimately
distinguish an exceptionally small lightest eigenvalue.

\begin{figure}[!ht]
\centering
\includegraphics[width=.6\textwidth]{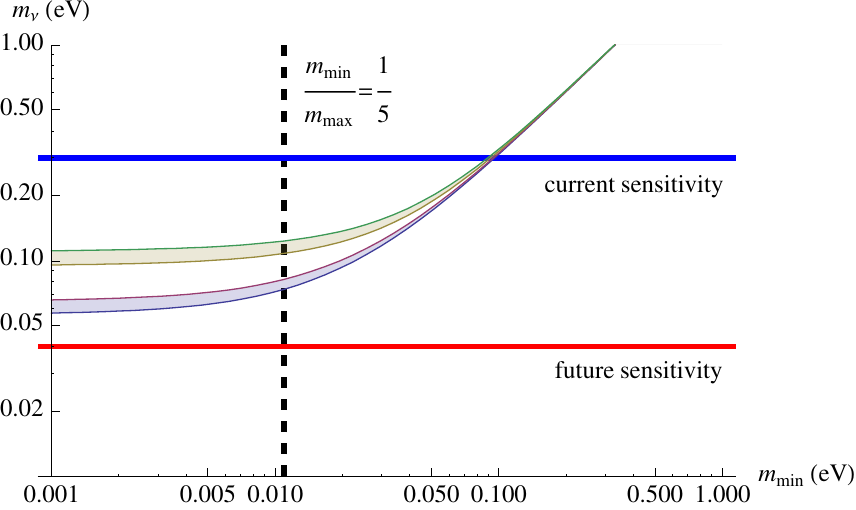}
\caption{Values of the sum of neutrino masses $m_\nu$ versus the smallest eigenvalue, $m_\mathrm{min}$, for either the normal hierarchy (blue) or inverted hierarchy (brown).  Conservatively, our model restricts us to the region $\frac{m_\mathrm{min}}{m_\mathrm{max}}\lesssim \frac 1 {5}$ (left of the dotted line), and thus gives a reasonably clear prediction for $m_\nu$.  For either the normal or inverted hierarchy, these predictions can't be tested with current precision but should be within reach of future studies \cite{Hannestad:2002cn}.}
\label{cosmologicalbound}
\end{figure}

Cosmological bounds might then be the best way to detect a small third eigenvalue in the case of a normal hierarchy, since we'd expect $m_\nu\sim \sqrt{\De m_{32}^2}\sim 0.05\eV$ (Fig. \ref{cosmologicalbound}), which could be within reach of future studies \cite{Hannestad:2002cn}.

In the case of the inverted hierarchy, the two larger masses are
both in the range $.046\eV<m<.056\eV$.
We then find $m_{\b\b}$
ranges from \be 0.013\eV< |m_{\b\b}|<.056\eV \ee with most of the
uncertainty from $\a_{12}$.  The correction from $m_3$ would be at
most about $\sin^2(\theta_{13}) m_1/10<0.003 \eV$. In this case  the
leading correction from a larger value for the smallest eigenvalue
than the $\alpha_\GUT m_1$ expected in our model would come from larger values
for both  $m_1$ and $m_2$, since only mass differences are known.
So a substantially    larger value for $m_{\b\b}$ than the minimal
value could rule out this model. However, with a measurement in the
above range it will be difficult to determine if the smallest
eigenvalue is bigger than we would expect in our model. The value,
though consistent with an inverted hierarchy, would not necessarily
be precise enough to determine the overall mass scale with
sufficient accuracy to pin down the mass of the lightest eigenvalue,
especially until the phase $\a_{12}$ is known.

Once again, cosmological bounds could test our model, since an inverted hierarchy with a small
third eigenvalue predicts $m_\nu\sim 2\sqrt{\De m_{32}^2}\sim 0.1\eV$, in reach of future studies (Fig. \ref{cosmologicalbound}).

\begin{figure}[!ht]
\centering
\includegraphics[width=.6\textwidth]{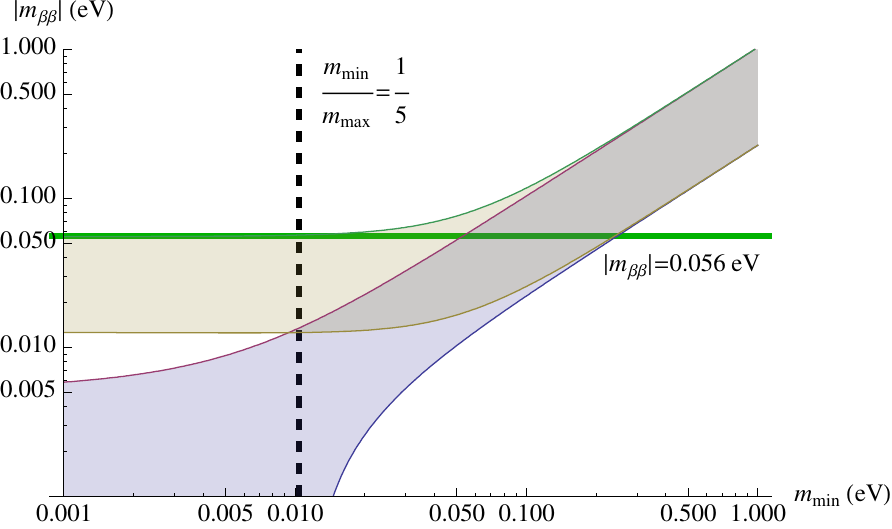}
\caption{Possible values of $|m_{\b\b}|$ versus the smallest mass eigenvalue $m_\mathrm{min}$.  The brown
shaded region corresponds to the inverted hierarchy, while the blue corresponds to the normal hierarchy.  The
uncertainty is from a combination of Majorana phases and uncertainty in the known values of $\De m^2_{12}, \De m^2_{23}$, and the mixing angles.  We predict $\frac{m_\mathrm{min}}
{m_{\mathrm{max}}}\aeq \a_\GUT \aeq \frac 1 {25}$, so a conservative upper bound for $m_\mathrm{min}$ would be $\frac{m_\mathrm{min}}{m_\mathrm{max}}=\frac 1 5$, about 5 times larger allowing for unknown order unity factors.  This yields a rough upper bound of $|m_{\b\b}|<0.056\eV$ (green line, above). A higher measured value for $|m_{\b\b}|$ would rule out our model.}
\label{doublebetadecaybounds}
\end{figure}

It's also clear that these models can be ruled out by planned neutrinoless double $\b$ decay
experiments. The largest possible value for $m_{\b\b}$ that we predict
is $.056 \eV$. Any larger value would indicate the overall scale of the
neutrino masses is bigger than would be expected from this rank-2
matrix form (Figure \ref{doublebetadecaybounds}).

\section{Conclusions} Given the many potential routes for getting from string
theory to the Standard Model, it is worthwhile to investigate
qualitatively new models that might give novel
insights into perplexing puzzles such as the hierarchy or flavor
problems. In some cases the models merely implement known mechanisms,
but sometimes they introduce genuinely new ideas. In other cases,
such as this one, the theory falls somewhat in between. Technically
the models we describe don't necessarily have mechanisms that cannot
be accounted for with symmetries and additional heavy fields.
Nonetheless the models from an effective theory viewpoint might be
cumbersome or somewhat artificial.

In this work we have shown the underlying mechanism that allows
F-theory models to reproduce the flavor structure of the Standard
Model. We have also shown how to obtain the correct pattern of
masses and mixings in the neutrino sector. We have furthermore shown
that an interesting prediction seems to be a dominantly rank-2
matrix for the neutrinos and we have shown how this prediction can
give testable consequences.

\section{Acknowledgements} We'd like to thank Matthew Buckley, Clay Cordova, Jonathan Heckman, Joe Marsano, Yasinori Nomura,
Sakura Schafer-Nameki, John Preskill, Jihye Seo, Sean Tulin, Mark Wise, and Cumrun Vafa
for useful comments on this work. LR thanks the California Institute
of Technology and the Moore Fellowship Program for their hospitality
while this work was completed. This research was supported in part
by NSF grant PHY-055611.

\nocite{*}                   
\bibliography{FtheoryFlavor}        

\providecommand{\href}[2]{#2}\begingroup\raggedright\begin{thebibliography}{10}

\bibitem{Grossman:1999ra}
Y.~Grossman and M.~Neubert, ``{Neutrino masses and mixings in non-factorizable
  geometry},'' \href{http://dx.doi.org/10.1016/S0370-2693(00)00054-X}{{\em
  Phys. Lett.} {\bf B474} (2000)  361--371},
\href{http://arxiv.org/abs/hep-ph/9912408}{{\tt arXiv:hep-ph/9912408}}.

\bibitem{Gherghetta:2000qt}
T.~Gherghetta and A.~Pomarol, ``{Bulk fields and supersymmetry in a slice of
  AdS},'' \href{http://dx.doi.org/10.1016/S0550-3213(00)00392-8}{{\em Nucl.
  Phys.} {\bf B586} (2000)  141--162},
\href{http://arxiv.org/abs/hep-ph/0003129}{{\tt arXiv:hep-ph/0003129}}.

\bibitem{Fitzpatrick:2007sa}
A.~L. Fitzpatrick, G.~Perez, and L.~Randall, ``{Flavor from Minimal Flavor
  Violation \& a Viable Randall- Sundrum Model},''
\href{http://arxiv.org/abs/0710.1869}{{\tt arXiv:0710.1869 [hep-ph]}}.

\bibitem{Perez:2008ee}
G.~Perez and L.~Randall, ``{Natural Neutrino Masses and Mixings from Warped
  Geometry},'' \href{http://dx.doi.org/10.1088/1126-6708/2009/01/077}{{\em
  JHEP} {\bf 01} (2009)  077},
\href{http://arxiv.org/abs/0805.4652}{{\tt arXiv:0805.4652 [hep-ph]}}.

\bibitem{Heckman:2008qa}
J.~J. Heckman and C.~Vafa, ``{Flavor Hierarchy From F-theory},''
\href{http://arxiv.org/abs/0811.2417}{{\tt arXiv:0811.2417 [hep-th]}}.

\bibitem{Beasley:2008dc}
C.~Beasley, J.~J. Heckman, and C.~Vafa, ``{GUTs and Exceptional Branes in
  F-theory - I},'' \href{http://dx.doi.org/10.1088/1126-6708/2009/01/058}{{\em
  JHEP} {\bf 01} (2009)  058},
\href{http://arxiv.org/abs/0802.3391}{{\tt arXiv:0802.3391 [hep-th]}}.

\bibitem{Beasley:2008kw}
C.~Beasley, J.~J. Heckman, and C.~Vafa, ``{GUTs and Exceptional Branes in
  F-theory - II: Experimental Predictions},''
  \href{http://dx.doi.org/10.1088/1126-6708/2009/01/059}{{\em JHEP} {\bf 01}
  (2009)  059},
\href{http://arxiv.org/abs/0806.0102}{{\tt arXiv:0806.0102 [hep-th]}}.

\bibitem{FutureMarsano}
M.~Liu, J.~Marsano, N.~Saulina, and S.~Schafer-Nameki, ``{In Progress},''.

\bibitem{Katz:1996xe}
S.~H. Katz and C.~Vafa, ``{Matter from geometry},''
  \href{http://dx.doi.org/10.1016/S0550-3213(97)00280-0}{{\em Nucl. Phys.} {\bf
  B497} (1997)  146--154},
\href{http://arxiv.org/abs/hep-th/9606086}{{\tt arXiv:hep-th/9606086}}.

\bibitem{Froggatt:1978nt}
C.~D. Froggatt and H.~B. Nielsen, ``{Hierarchy of Quark Masses, Cabibbo Angles
  and CP Violation},''
\href{http://dx.doi.org/10.1016/0550-3213(79)90316-X}{{\em Nucl. Phys.} {\bf
  B147} (1979)  277}.

\bibitem{Amsler:2008zzb}
{\bf Particle Data Group} Collaboration, C.~Amsler {\em et al.}, ``{Review of
  particle physics},''
\href{http://dx.doi.org/10.1016/j.physletb.2008.07.018}{{\em Phys. Lett.} {\bf
  B667} (2008)  1}.

\bibitem{SanchezTalk}
{\bf MINOS} Collaboration, M.~Sanchez, ``{Initial results for $\nu_\mu\to\nu_e$
  oscillations in MINOS},'' {\em Talk given at FNAL seminar} (2009)  .

\bibitem{Heckman:2008es}
J.~J. Heckman, J.~Marsano, N.~Saulina, S.~Schafer-Nameki, and C.~Vafa,
  ``{Instantons and SUSY breaking in F-theory},''
\href{http://arxiv.org/abs/0808.1286}{{\tt arXiv:0808.1286 [hep-th]}}.

\bibitem{Marsano:2008py}
J.~Marsano, N.~Saulina, and S.~Schafer-Nameki, ``{An Instanton Toolbox for
  F-Theory Model Building},''
\href{http://arxiv.org/abs/0808.2450}{{\tt arXiv:0808.2450 [hep-th]}}.

\bibitem{FutureHV}
V.~Bouchard, J.~J. Heckman, J.~Seo, and C.~Vafa, ``{In Preparation},''.

\bibitem{Kaplinghat:2003bh}
M.~Kaplinghat, L.~Knox, and Y.-S. Song, ``{Determining neutrino mass from the
  CMB alone},'' \href{http://dx.doi.org/10.1103/PhysRevLett.91.241301}{{\em
  Phys. Rev. Lett.} {\bf 91} (2003)  241301},
\href{http://arxiv.org/abs/astro-ph/0303344}{{\tt arXiv:astro-ph/0303344}}.

\bibitem{Hannestad:2002cn}
S.~Hannestad, ``{Can cosmology detect hierarchical neutrino masses?},''
  \href{http://dx.doi.org/10.1103/PhysRevD.67.085017}{{\em Phys. Rev.} {\bf
  D67} (2003)  085017},
\href{http://arxiv.org/abs/astro-ph/0211106}{{\tt arXiv:astro-ph/0211106}}.

\bibitem{Vogel:2006sq}
P.~Vogel, ``{Neutrinoless double beta decay},''
\href{http://arxiv.org/abs/hep-ph/0611243}{{\tt arXiv:hep-ph/0611243}}.

\bibitem{Heckman:2008qt}
J.~J. Heckman and C.~Vafa, ``{F-theory, GUTs, and the Weak Scale},''
\href{http://arxiv.org/abs/0809.1098}{{\tt arXiv:0809.1098 [hep-th]}}.

\bibitem{Heckman:2008rb}
J.~J. Heckman and C.~Vafa, ``{From F-theory GUTs to the LHC},''
\href{http://arxiv.org/abs/0809.3452}{{\tt arXiv:0809.3452 [hep-ph]}}.

\end{thebibliography}\endgroup
\bibliographystyle{utphys}   

\end{document}